\documentclass[fleqn,10pt]{wlscirep}
\usepackage[utf8]{inputenc}
\usepackage[T1]{fontenc}
\usepackage{babel}
\usepackage{amsmath}
\usepackage{graphicx}
\usepackage{multirow}
\usepackage{multicol}
\usepackage{fancyhdr}
\usepackage{hyphenat}
\usepackage{url}
\usepackage{hyperref}

\title{Mobile BCI dataset of scalp- and ear-EEGs with ERP and SSVEP paradigms while standing, walking, and running}

\author[1]{Young-Eun Lee}
\author[1]{Gi-Hwan Shin}
\author[1]{Minji Lee}
\author[1,2,*]{Seong-Whan Lee}
\affil[1]{Korea University, Department of Brain and Cognitive Engineering, Seoul, 02841, Republic of Korea}
\affil[2]{Korea University, Department of Artificial Intelligence, Seoul, 02841, Republic of Korea}

\affil[*]{corresponding author: Seong-Whan Lee (sw.lee@korea.ac.kr)}

\begin{abstract}
We present a mobile dataset obtained from electroencephalography (EEG) of the scalp and around the ear as well as from locomotion sensors by 24 participants moving at four different speeds while performing two brain-computer interface (BCI) tasks. 
The data were collected from 32-channel scalp-EEG, 14-channel ear-EEG, 4-channel electrooculography, and 9-channel inertial measurement units placed at the forehead, left ankle, and right ankle.
The recording conditions were as follows: standing, slow walking, fast walking, and slight running at speeds of 0, 0.8, 1.6, and 2.0\,m/s, respectively. 
For each speed, two different BCI paradigms, event-related potential and steady-state visual evoked potential, were recorded. 
To evaluate the signal quality, scalp- and ear-EEG data were qualitatively and quantitatively validated during each speed. We believe that the dataset will facilitate BCIs in diverse mobile environments to analyze brain activities and evaluate the performance quantitatively for expanding the use of practical BCIs.
\end{abstract}
\begin{document}

\flushbottom
\maketitle
figshare DOI: https://doi.org/10.6084/m9.figshare.16669072

\thispagestyle{empty}


\section*{Background \& Summary}

Human movement is a complex process that requires the integration of the central and peripheral nervous systems, and therefore, researchers have analyzed human locomotion using brain activity \cite{cite1,cite2}.
Brain-computer interface (BCI) has been studied based on the communication between human thoughts and external devices to recover the motor sensory function of disabled patients and support daily life of healthy people. \cite{cite3,cite4,cite5}. 
In particular, electroencephalography (EEG) has been used as the most common method for measuring brain activity with high time resolution, portability, and ease of use \cite{cite6}; in addition, several attempts have been made to increase its practicality \cite{cite7,cite8,cite9,cite10}.
However, EEG recording in a mobile environment can cause artifacts and signal distortion, further resulting in loss of accuracy and signal quality \cite{cite11,cite12}.
Therefore, research in the mobile environment is necessary to study the brain activity during movements to mitigate limitations such as loss of accuracy and signal quality and to improve the practical BCI technology \cite{cite9,cite12}.
Moreover, several studies have developed software techniques to realize the practical BCI, such as preprocessing algorithm removing artifacts \cite{cite13,cite14,cite15,cite16} or novel classification algorithm improving user intention performance \cite{cite12,cite17}.

To recognize the human intention, two representative exogenous BCI paradigms, event-related potential (ERP) \cite{cite18} and steady-state visual evoked potential (SSVEP) \cite{cite19}, are commonly used in the mobile environment owing to their strong responses to brain activity. 
ERP is a time-locked brain response to stimuli (i.e., visual, auditory, etc.), including a positive peak response (P300) that occurs 300\,ms after the stimulus appears. The ERP has a relatively high performance in both scalp-EEG and ear-EEG, with accuracies of 85--95\% for scalp-EEG \cite{cite20,cite21} and approximately 70\% for ear-EEG \cite{cite9} in a static state.
SSVEP is a period brain response in the occipital area to stimuli flickering at a particular frequency. The performance of SSVEP is reliable in terms of the accuracy and signal-to-noise ratio (SNR), with 80--95\% accuracy for scalp-EEG \cite{cite17,cite20}, but 40--70\% accuracy for ear-EEG as it is located far from the occipital cortex \cite{cite22,cite23}.
Brain signal data obtained by performing BCI paradigms can be used to quantitatively evaluate signal quality in a mobile environment \cite{cite24}.

Portable and non-hair-bearing EEG have been frequently investigated to enhance the applicability of practical BCI in the real world \cite{cite25,cite26,cite27,cite28,cite29}. In particular, ear-EEG, which comprises electrodes placed inside or around the ear, has several advantages over conventional scalp-EEG in terms of stability, portability, and unobtrusiveness \cite{cite9, cite30}. Moreover, the signal quality of ear-EEG has been validated for recognizing human intention using several BCI paradigms, including ERP \cite{cite9,cite29,cite31}, SSVEP \cite{cite22,cite23}, and others \cite{cite32}. 


Recently, EEG datasets for mobile environments have been published, including motion information and different mobile environments.
He et al. \cite{cite33} recorded signals from 60-channel scalp-EEG, 4-channel electrooculography (EOG), and 6 goniometers from 8 participants while walking slowly at a constant speed of 0.45\,m/s. They used a BCI paradigm, avatar control, by predicting the joint angle of goniometers from the EEG while walking.
Brantley et al. \cite{cite34} collected a dataset consisting of full-body locomotion from 10 participants on stairs, ramps, and level grounds without BCI paradigms, by recording from 60-channel scalp-EEG, 4-channel EOG, 12-channel electromyogram (EMG), and 17 inertial measurement units (IMUs).
Wagner et al. \cite{cite35} recorded signals from 108-channel scalp-EEG, 2-channel EMG, 2 pressure sensors, and 3 goniometers from 20 participants without BCI paradigms while walking at a constant speed.
Consequently, although these datasets were collected from the mobile environment, the movement condition was kept constant, and two of the datasets were performed without BCI paradigms. In addition, because only scalp-EEG signals were measured, the application of practical BCI was restricted.

In this study, we present a mobile BCI dataset with scalp- and ear-EEGs collected from 24 participants with BCI paradigms at different speeds. Data from 32-channel scalp-EEG, 14-channel ear-EEG, 4-channel EOG, and 27-channel IMUs were recorded simultaneously. The experimental environment involved movements of participants at different speeds of 0, 0.8, 1.6, and 2.0\,m/s on a treadmill. For each speed, two BCI paradigms were used to evaluate signal quality, which facilitated diverse analysis, including time-domain analysis using ERP data and frequency-domain analysis using SSVEP data. 
Therefore, we believe that the dataset facilitates addressing the issues of brain dynamics in diverse mobile environments in terms of the cognitive level of multitasking, locomotion complexity, and quantitative evaluation of artifact removal methods or classifiers for BCI tasks in the mobile environment.

\section*{Methods}
\subsection*{Participants}
Twenty-four healthy individuals (14 men and10 women, 24.5 $\pm$ 2.9 years of age) without any history of neurological or lower limb pathology participated in this experiment. In the ERP tasks, all of them participated, and 17 participants performed a slight running session at a speed of 2.0\,m/s. In the SSVEP tasks, 23 of them participated, excluding one because of a personal problem unrelated to the experimental procedure; 16 participants performed a slight running session at a speed of 2.0\,m/s. The participants were provided the option to perform a slight running session. This study was approved by the Institutional Review Board of Korea University (KUIRB-2019-0194-01), and all participants provided written informed consent before the experiments. All experiments were conducted in accordance with the Declaration of Helsinki.

\subsection*{Data Acquisition} 
\begin{figure*}[t!]
\centering
\scriptsize
\centerline{\includegraphics[width=\textwidth]{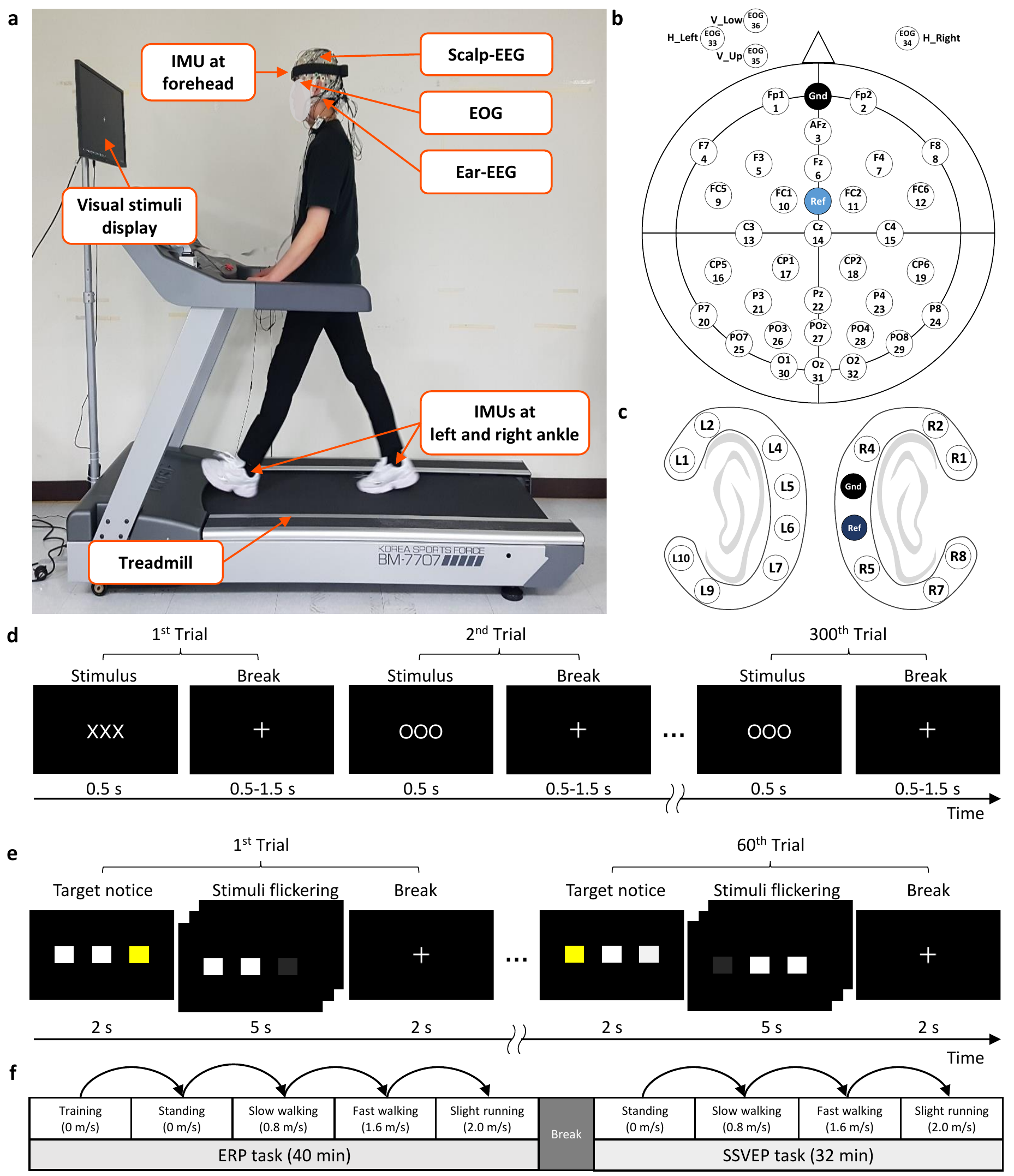}}
\caption{Experimental design. (a) Experimental setup while standing (0\,m/s), slow walking (0.8\,m/s), fast walking (1.6\,m/s), and slight running (2.0\,m/s) on the treadmill, wearing scalp-EEG, ear-EEG, EOG, and IMUs. Informed consent was obtained from the participant for publishing the figure. Channel placement of (b) scalp-EEG with EOG and (c) ear-EEG. Experimental paradigms for (d) ERP paradigm with 300 trials and (e) SSVEP paradigm with 60 trials. (f) Experimental procedure.}
\label{fig:fig1}
\end{figure*}

For conducting the experiment, we simultaneously collected data from three different modalities: scalp-EEG, ear-EEG, and IMU (Figure \ref{fig:fig1}a, b, and c). All data can be accessed from here \cite{cite36}. To synchronize the three devices, triggers were sent to the recording system of each device simultaneously while presenting a paradigm in MATLAB.

The head circumference of each participant was measured to select an appropriately sized cap of scalp-EEG. We obtained signals from the scalp with 32 EEG Ag/AgCl electrodes according to the 10/20 international system using BrainAmp (Brain Product GmbH). The ground and reference electrodes were placed on Fpz and FCz. In addition, we used four EOG channels to capture dynamically changing eye movements such as blinking. The EOG channels were placed above and below the left eye to measure the vertical eye artifacts (VEOG$_{U}$ and VEOG$_{L}$) as well as at the left and right temples to measure the horizontal eye artifacts (HEOG$_{L}$ and HEOG$_{R}$) (Figure \ref{fig:fig1}b). The sampling rate of the scalp-EEG and EOG was 500\,Hz with a resolution of 32 bits. All electrode impedances were maintained below 50 $k\Omega$, and most channels were reduced to below 20 $k\Omega$ \cite{cite33,cite34}.

The ear-EEG consists of cEEGrids electrodes located around each ear of the participant, with eight channels on the left, six channels on the right, and ground and reference channels on the right at the center (Figure \ref{fig:fig1}c) \cite{cite9}. Two cEEGrids were connected to a wireless mobile DC EEG amplifier (SMARTING, mBrainTrain, Belgrade, Serbia). Data were recorded at a sampling rate of 500\,Hz and a resolution of 24 bits. All impedances of ear-EEG were maintained below 50 $k\Omega$, and most channels were reduced to below 20 $k\Omega$ \cite{cite33,cite34}.

To measure the locomotion of the participants, three wearable IMU sensors (APDM wearable technologies) were placed at the head, left ankle, and right ankle. An IMU consisted of 9-channel sensors, including a 3-axis accelerometer, 3-axis gyroscope, and 3-axis magnetometer. Therefore, 27-channel IMU signals were collected and recorded at a sampling rate of 128 Hz with a resolution of 32 bits.

\subsection*{Experimental Paradigm}
They performed tasks under the ERP and SSVEP paradigms, in which stimuli were displayed on the monitor during each session at different speeds. Two BCI paradigms were developed based on the OpenBMI (http://openbmi.org) \cite{cite20} and Psychtoolbox (http://psychtoolbox.org) \cite{cite37} in MATLAB (The Mathworks, Natick, MA).

During the ERP task, target (`OOO') and non-target (`XXX') characters were presented on the monitor as visual stimuli. All characters were displayed with a black background at the center of the monitor screen. The proportion of the target was 0.2, and the total number of trials was 300. In a trial, one of the stimuli was presented for 0.5\,s, and a fixation cross (`+') was presented to take a break for randomly 0.5--1.5\,s (Figure \ref{fig:fig1}d).

During the SSVEP task, three target SSVEP visual stimuli were displayed at three positions (left, center, and right) on an LCD monitor \cite{cite19,cite38}.
The frequency range of stimuli containing 5--30\,Hz is known to be appropriate for obtaining SSVEP responses \cite{cite39}. It is also known that movement artifacts have a significant impact on frequency spectrum below 12\,Hz \cite{cite40}. Based on these studies, the stimuli were designed to flicker at 5.45, 8.57, and 12\,Hz, which were calculated by dividing the monitor frame rate of 60\,Hz by an integer (i.e., 60/11, 60/7, and 60/5) \cite{cite20}.
The participants were asked to gaze in the direction of the target stimulus highlighted in yellow. In each trial, the target of a random sequence was noticed for 2\,s, after which all stimuli blinked for 5\,s, with a break time for 2\,s. The SSVEP experiment consisted of 20 trials for each frequency, a total of 60 trials in a session (Figure \ref{fig:fig1}e).

\subsection*{Experimental Protocol and Procedure} 
Figure \ref{fig:fig1}a depicts the experimental setup of this study. The participants stood on the treadmill in a lab and were instructed to look at a 24-inch monitor (refresh rate: 60 Hz, resolution: 1920 $\times$ 1080 pixels) placed 80 ($\pm$ 10) cm in front of the participants. The participants were monitored and instructed to minimize other movements, such as that of neck or arm, to avoid any artifacts that might occur by movements other than walking.
The mobile environment included standing, slow walking, fast walking, and slight running at speeds of 0, 0.8, 1.6, and 2.0\,m/s, respectively on the treadmill (0$^{\circ}$ inclination) \cite{cite41,cite42}. 

To proceed with the experiment under the same conditions for each participant, the experimental procedures were sequentially performed (Figure \ref{fig:fig1}f).
They conducted two BCI tasks while standing, slow walking, fast walking, and slight running on the treadmill. Training sessions for the ERP task were conducted at a speed of 0\,m/s to train the ERP classifier prior to all ERP tasks. Duration of a session of ERP and SSVEP tasks consisted of 7--8 min, with a total of 40 min and 32 min for all sessions. Each session involved the same procedure with a random sequence of targets. All sessions for one participant were performed on a single day.
Moreover, due to the possibility of fatigue and habitation, which could be induced in the sequence of the experiment \cite{cite43}, the following actions were considered.
At first, participants were allowed sufficient breaks between and within sessions when needed.
Furthermore, the participants became familiar with the paradigm stimuli by being fully exposed to them before starting the experiment.

\subsection*{Preprocessing}
We preprocessed the data using an open-source toolbox for EEG data, such as BBCI (https://github.com/bbci/bbci\_public) \cite{cite44}, BCILAB
(https://github.com/sccn/BCILAB) \cite{cite45}, and EEGLAB (https://sccn.ucsd.edu/eeglab) \cite{cite46} in MATLAB.
At first, the data were preprocessed using a high-pass filter that was set above 0.5\,Hz using a fifth-order Butterworth filter. 
Thereafter, three procedures: EOG removal, line noise removal, and interpolation were performed. Vertical EOG components were removed from the scalp-EEG using the \textit{flt\_eog} function in BCILAB \cite{cite47}. 
The line noise removal method automatically removed artifacts that contained noise for extended periods of time with several parameters.
This method removed bad channels that carried abnormal signals with standard deviations above the threshold of z-score using the function \textit{flt\_clean\_channels} in BCILAB with threshold of 4 and window length of 5\,s.
The removed bad channels were interpolated using the super-fast spherical method to avoid losing any channel information. On average, 2.38 $\pm$ 1.94 channels in the scalp-EEG and 1.35 $\pm$ 1.18 channels in the ear-EEG were removed and interpolated for all participants in all sessions. 
All channels in scalp-EEG and ear-EEG were each re-referenced to a common average reference. 
We down-sampled the data from scalp-EEG, ear-EEG, and IMU sensors to 100\,Hz.
The continuous signal was segmented into epoched signals according to the time of each paradigm. For the ERP, each trial was segmented from -200--800\,ms based on the stimulus presentation time. For the SSVEP, each trial was segmented from 0--5\,s based on the starting time of the stimulus flickering.

\section*{Data Records}
All data files are available in Open Science Framework repository \cite{cite36} and are available under the terms of Attribution 4.0 International Creative Commons License (http://creativecommons.org/licenses/by/4.0/).

\subsection*{Data Format}
\begin{figure*}[t!]

\centering
\scriptsize
\centerline{\includegraphics[width=\textwidth]{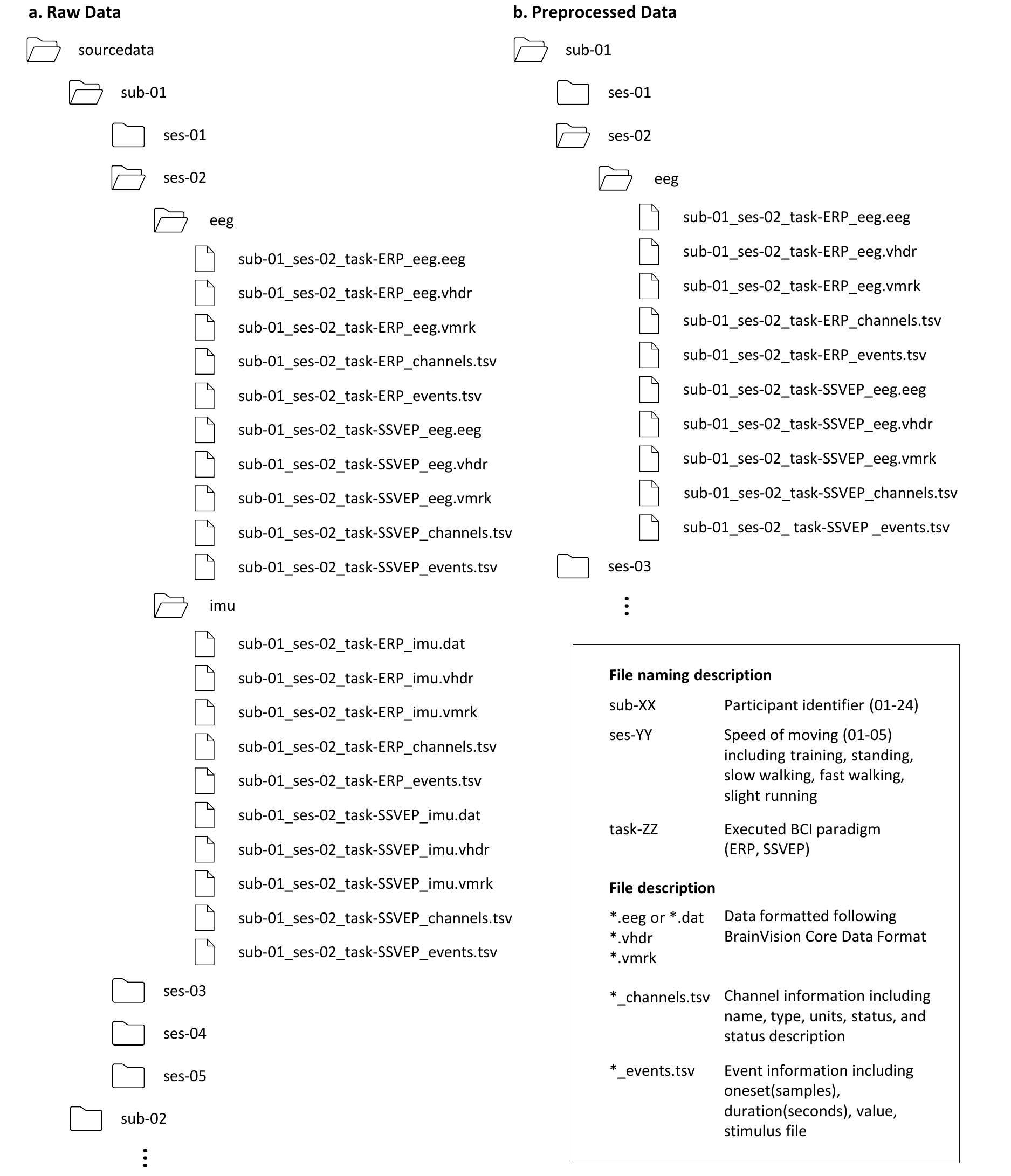}}
\caption{Data folder structure. The folders and files are described, including (a) raw data and (b) preprocessed data in the data repository. The folder was named with `sub-XX', `ses-YY', and modality, and the file was named with `sub-XX\_ses-YY\\\_task-ZZ\_WW'. The `sub-XX' indicated the participant identifier, including 1--24, the `ses-YY' indicated the session number, including training(01), standing(02), slow walking(03), fast walking(04), and slight running(05), the `task-ZZ' indicated executed BCI paradigms including ERP and SSVEP, and the modality `WW' indicated the modality of each data, including EEG (scalp-EEG and ear-EEG) and IMU.}

\label{fig:fig2}

\end{figure*}

All data are provided according to the standardized Brain Imaging Data Structure format for EEG data \cite{cite48} as shown in Figure \ref{fig:fig2}. The data format followed BrainVision Core Data Format, developed by Brain Products GmbH. The data file is organized with the following naming convention:

\begin{center} sub-XX\_ses-YY\_task-ZZ\_eeg \end{center}

\noindent
where the session number includes 1--5, which session 1 indicates training session for ERP and session 2--5 indicates each speed of 0, 0.8, 1.6, and 2.0\,m/s, respectively, and the task includes ERP and SSVEP.
In `sourcedata' folder, the data are separated into EEG and IMU because their sampling frequencies are different. In each subject folder, the sampling frequencies of data are downed to 100\,Hz and data from all modalities are in a file.
The number of channels for each modality is listed in Table \ref{tab:tab1}.

\subsection*{Missing Data}
\subsubsection*{Data Missing}
The IMU data of participant 21 for ERP at 0\,m/s and that of participant 12 for SSVEP at 0.8\,m/s were missing because of a malfunction in the communication of the IMU during data collection.
\subsubsection*{Trials Missing}
The number of trials for ERP data of participant 11 at every speed and participant 13 and 15 at a speed of 2.0\,m/s, and SSVEP data of participant 14 at a speed of 2.0\,m/s were approximately two-thirds of the normal number of trials because of the malfunction of device communication.
\subsubsection*{Excluded Data}
The data of participant 17 at 2.0\,m/s for SSVEP, participant 19 at 2.0\,m/s for ERP and SSVEP, and participant 20 at 2.0\,m/s for ERP were excluded since the electrodes did not adhere well during data recording, resulting in loss of more than 50\% in a session.

\section*{Technical Validation}

\begin{figure*}[t!]
\centering
\scriptsize
\centerline{\includegraphics[width=\textwidth]{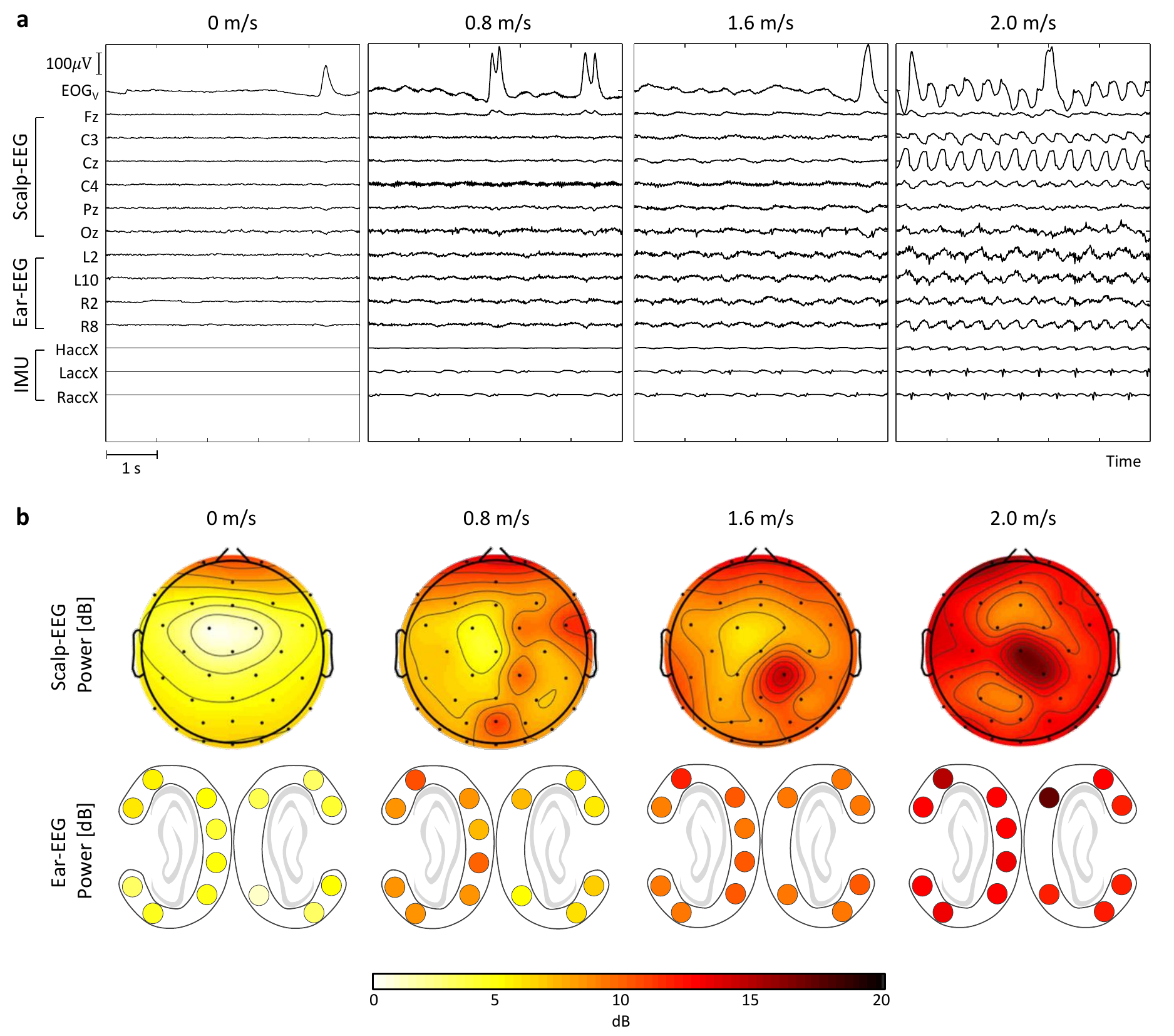}}
\caption{Examples of the signals and topography at different speeds. (a) Time-synchronized subset of scalp-EEG, ear-EEG, and IMU data for 5\,s while moving at different speeds of 0, 0.8, 1.6, and 2.0\,m/s. The $EOG_V$ channel was calculated by subtracting lower VEOG from upper VEOG. (b) EEG power topography in each channel of scalp-EEG and ear-EEG.}
\label{fig:fig3}
\end{figure*}

We evaluated the preprocessed data using the ERP and SSVEP paradigms in terms of accuracy and SNR.
Statistical analysis was conducted using a one-tailed paired t-test to compare the performance at each moving speed to the performance at standing, as indicated by the asterisk at a confidence level of 95\%.
Moreover, the ERP waves and power spectral density (PSD) for SSVEP were used to evaluate the signal quality.
Figure \ref{fig:fig3}a depicts an example of the scalp-EEG, ear-EEG, and IMU signals for 5\,s at speeds of 0, 0,8, 1.6, and 2.0\,m/s. The amplitudes of the scalp-EEG, ear-EEG, and IMU increased as the speed increased \cite{cite41, cite49, cite50}. Figure \ref{fig:fig3}b depicts an example of the topography for the scalp-EEG and ear-EEG at different speeds of 0, 0.8, 1.6, and 2.0\,m/s. The powers of the scalp-EEG and ear-EEG increased as the speed increased.

\subsection*{Statistical Verification}
\begin{figure*}[t!]
\centering
\scriptsize
\centerline{\includegraphics[width=\textwidth]{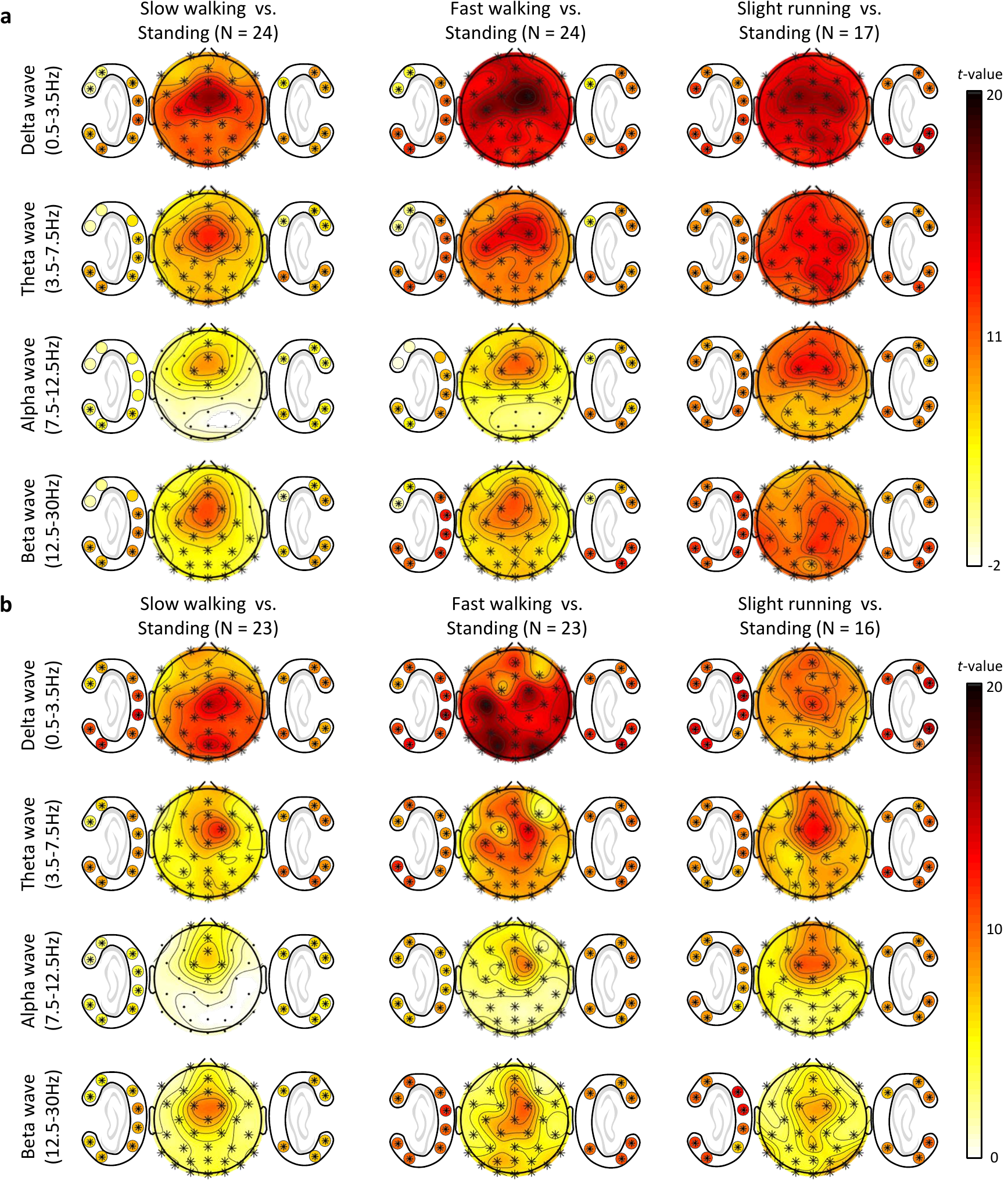}}
\caption{Statistical differences of PSD in each frequency band for scalp- and ear-EEGs between standing and other speeds while (a) ERP and (b) SSVEP. The colored topological maps indicate t-values and the electrodes in cluster showing a statistically significant effect on spectral power between the data of corresponding speeds are marked with black asterisk (\textit{p} $<$ 0.05, cluster-based correction for multiple comparison).}
\label{fig:fig4}
\end{figure*}

To verify the dataset, we performed statistical verification to demonstrate significant differences between the speeds across every channel of scalp-EEG and ear-EEG. Figure \ref{fig:fig4}a for the ERP and Figure \ref{fig:fig4}b for the SSVEP depicts the topological map of t-values in particular frequency bands, including delta waves (0.5--3.5\,Hz), theta waves (3.5--7.5\,Hz), alpha waves (7.5--12.5\,Hz), and beta waves (12.5-30\,Hz) \cite{cite51}.
In particular, PSDs in each frequency band were analyzed using cluster-based correction with non-parametric permutation testing for multiple comparisons to verify the difference between the data at four speeds, including 0, 0.8, 1.6, and 2.0\,m/s. The significance probabilities and critical values of permutation distribution are estimated using Monte-Carlo method with iterations of 10,000.

Significant channels could indicate that noise signals are included in corresponding frequency bands and speeds. The topography of the delta band depicts that step frequencies, which was mostly in the range of 0.5--3.5\,Hz, affect most channels at all speeds. 
During slight running session, all channels, including scalp-EEG and ear-EEG, were significantly different in entire frequency band.
In addition, paradigm-related areas such as the occipital area during SSVEP tasks and the central area during ERP tasks showed the large \textit{t}-values in delta band, resulting in low concentration on the tasks due to the workload of multi-tasking.

\subsection*{Evaluation of ERP}
The ERP dataset was evaluated by demonstrating ERP waves and metrics using the area under the receiver operating characteristic curve (AUC) and approximate SNR at each speed. AUC indicates the true positive rate over the false positive rate of the results. To acquire the AUC, the features of the ERP were extracted by the power over time intervals of every 50\,ms from 200\,ms to 450\,ms. For the classification, we used a conventional classifier, regularized linear discriminant analysis, to evaluate the ERP performance. The data from the training session at a speed of 0\,m/s were used for the training set, and the other dataset containing different speeds was used for the testing set.
The SNR can indicate the quality of signals, and approximate SNR of ERP was calculated by the root mean square (RMS) of the amplitude of the peaks at P300 divided by the RMS of the average amplitude of the pre-stimulus baseline (-200--0\,ms) at channel Pz \cite{cite52,cite53}.

\begin{figure*}[t!]
\centering
\scriptsize
\centerline{\includegraphics[width=\textwidth]{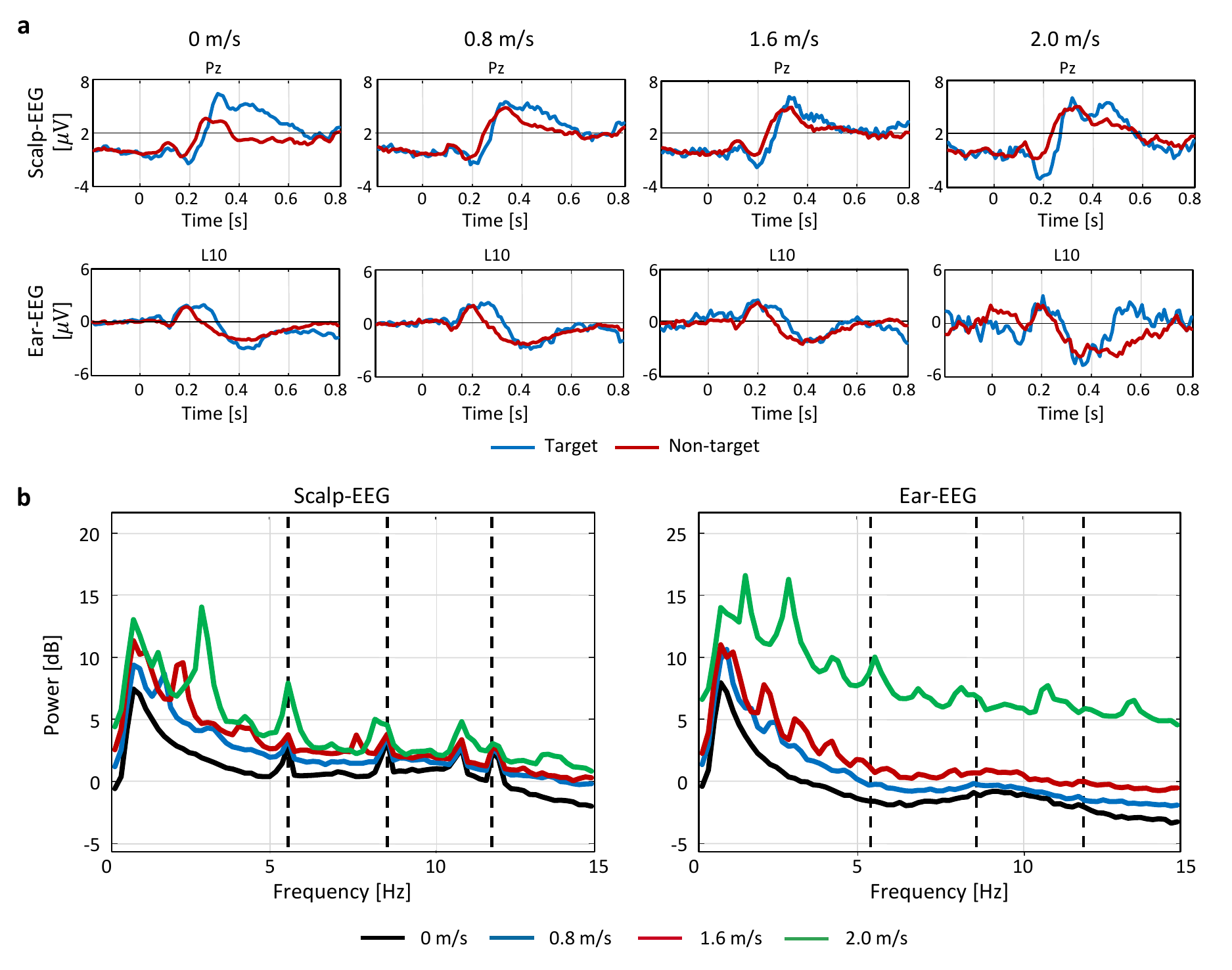}}
\caption{Grand average of all participants of ERP and SSVEP waveforms according to four different speeds of 0, 0.8, 1.6, and 2.0\,m/s. (a) Grand average of all participants of ERP waves of target and non-target stimuli in scalp-EEG and ear-EEG for 1\,s from -200--800\,ms according to the trigger at different speeds. (b) Grand average PSD of all participants for SSVEP in scalp-EEG (left) and ear-EEG (right) at different speeds. The dash line indicated the target frequency, such as 5.45, 8.57, and 12\,Hz.}
\label{fig:fig5}
\end{figure*}

Figure \ref{fig:fig5}a depicts the ERP waves of the target and non-target stimuli in the scalp- and ear-EEGs at channels Pz and L10 at each speed. The higher the speed, the lower the amplitude of the P300 components of the target in both the scalp- and ear-EEGs. 
Tables \ref{tab:tab2} and \ref{tab:tab3} list the performance of ERP in the scalp-EEG and ear-EEG, respectively. The grand average AUCs of ERP for all participants were 0.90 $\pm$ 0.07, and 0.67 $\pm$ 0.07 $(p<0.05)$ in the scalp-EEG at speeds of 0 and 1.6\,m/s, respectively, and 0.72 $\pm$ 0.14 and 0.58 $\pm$ 0.06 $(p<0.05)$ in the ear-EEG at speeds of 0 and 1.6\,m/s, respectively. The grand average SNRs of ERP for all participants were 0.95 $\pm$ 0.09 and 1.06 $\pm$ 0.14 $(p<0.05)$ for the scalp-EEG at speeds of 0 and 1.6\,m/s, respectively, and 1.06 $\pm$ 0.27 and 0.98 $\pm$ 0.05 for the ear-EEG at speeds of 0 and 1.6\,m/s, respectively.

\subsection*{Evaluation of SSVEP}
The SSVEP dataset was evaluated by implementing statistical analysis to measure the signal properties using PSD, and the metrics using accuracy and approximate SNR at each speed. Accuracy was measured as the percentage of correct predictions in the total number of cases. A canonical correlation analysis was used for the classification that does not require the training datasets.
The SNR of SSVEP was calculated using the ratio of the power of the target frequencies to the power of the neighboring frequencies (resolution: 0.25\,Hz, number of neighbors: 12) \cite{cite54}.

Figure \ref{fig:fig5}b depicts the PSD of the SSVEP for the scalp-EEG and ear-EEG at channels Oz and L10 at each speed. The higher the speed, the greater the power in all frequency spectra for both the scalp- and ear-EEGs.
Tables \ref{tab:tab4} and \ref{tab:tab5} list the performance of the SSVEP for scalp-EEG and ear-EEG, respectivelyThe grand average accuracies of SSVEP for all participants were 88.70 $\pm$ 19.52\% and 80.65 $\pm$ 20.38\% $(p<0.05)$ for scalp-EEG at speeds of 0 and 1.6\,m/s, respectively, and 53.19 $\pm$ 13.93 and 39.57 $\pm$ 6.39 $(p<0.05)$ for ear-EEG at speeds of 0 and 1.6\,m/s, respectively. The grand average SNRs of SSVEP for all participants were 2.64 $\pm$ 0.99 and 1.92 $\pm$ 0.68 $(p<0.05)$ for the scalp-EEG at speeds of 0 and 1.6\,m/s, respectively, and 1.21 $\pm$ 0.23 and 1.03 $\pm$ 0.10 $(p<0.05)$ for the ear-EEG at speeds of 0 and 1.6\,m/s, respectively.

\section*{Usage Notes}
This mobile dataset is available in the BrainVision Core Data Format. For analyzing the dataset, we recommend using a common open-source toolbox for EEG data, such as BBCI (https://github.com/bbci/bbci\_public) \cite{cite44}, OpenBMI (http://openbmi.org) \cite{cite20}, and EEGLAB (https://sccn.ucsd.edu/eeglab) \cite{cite46} in the MATLAB environment, or MNE (https://martinos.org/mne) \cite{cite55} in the Python environment. The supporting code is available on GitHub (https://github.com/DeepBCI/Deep-BCI). For the preprocessing, we recommend performing down-sampling to give all signals equal sampling frequency, filtering out extremely low frequency below 0.1\,Hz at least to remove the DC drift using a high-pass filter, and interpolating the high distributed channels among all channels. This dataset can be used for the performance evaluation of artifact removal methods and analysis of mental states with quantitative evaluation via BCI paradigms in a mobile environment.

\section*{Code Availability}
The MATLAB scripts are available for loading data, for evaluating classification performance or signal quality, and for plotting figures at \href{https://github.com/youngeun1209/MobileBCI\_Data}{https://github.com/youngeun1209/MobileBCI\_Data}.

\section*{Acknowledgements} 

This work was supported by the Institute for Information \& Communications Technology Planning \& Evaluation (IITP) grant funded by the Korean Government, No. 2017-0-00451: Development of BCI based Brain and Cognitive Computing Technology for Recognizing User’s Intentions using Deep Learning, No. 2015-0-00185: Development of Intelligent Pattern Recognition Softwares for Ambulatory Brain-Computer Interface, and No. 2019-0-00079: Artificial Intelligence Graduate School Program (Korea University).
We would like to express our sincere gratitude to Y.-H. Kang and D.-Y. Lee for their assistance in data collection, and N.-S. Kwak for his advice while designing the experiment.

\section*{Author Contributions Statement}
Y.-E.L. contributed to the design of the experiment, data collection, software programming, data validation, and preparation of the manuscript. G.-H.S. contributed to data collection, software programming, data validation, and preparation of the manuscript. M.L. contributed to data validation and preparation of the manuscript. S.-W.L. contributed to supervision of the project and editing the manuscript.

\section*{Competing Interests} 
The authors declare no competing interests.

\section*{Figures \& Tables}


\newpage

\newpage


\clearpage

\newpage

\begin{table*}[t!]
\centering
\small
\renewcommand{\arraystretch}{1.3}
\renewcommand{\tabcolsep}{5mm}
\begin{tabular*}{\textwidth}{c|c|l}
\hline
\textbf{Modality} & \textbf{The   number of  channels} & \textbf{Description}  \\ \hline
Scalp-EEG     & 36     & 32   EEG channels and 4 EOG channels                                   \\ \hline
Ear-EEG       & 14     & 8   channels on the left side   and 6 channels on the right side    \\ \hline
IMU           & 27     & \begin{tabular}[c]{@{}l@{}} 3 devices (head, left ankle, and right ankle)\\ 9-axis device containing accelerometers, gyroscopes, and magnetometers\end{tabular} \\ \hline

\end{tabular*}
\caption{Description of channel types. Asterisk indicates significance levels of 1\% between the performance at 0\,m/s and corresponding speed.}
\label{tab:tab1}
\end{table*}

\begin{table*}[t!]
\centering
\small
\renewcommand{\arraystretch}{1.3}
\renewcommand{\tabcolsep}{5.05mm}
\begin{tabular*}{\textwidth}{c|c|c|c|c|c|c|c|c}
\hline
\textbf{Measure}      & \multicolumn{4}{c|}{\textbf{AUC}}   &   \multicolumn{4}{c}{\textbf{SNR}}    \\ \hline 
\textbf{Speed} & { 0 m/s } & {0.8 m/s} & {1.6 m/s} & {2.0 m/s} & { 0 m/s } & {0.8 m/s} & {1.6 m/s} & {2.0 m/s} \\ \hline 
\textbf{s1}  & 0.99                     & 0.83                     & 0.77                     & 0.59                     & 1.03                     & 1.01                     & 0.96                     & 0.99                     \\
\textbf{s2}  & 0.85                     & 0.68                     & 0.58                     & 0.61                     & 0.89                     & 0.91                     & 0.95                     & 0.97                     \\
\textbf{s3}  & 0.78                     & 0.76                     & 0.68                     & 0.56                     & 0.77                     & 0.83                     & 0.88                     & 0.96                     \\
\textbf{s4}  & 0.86                     & 0.84                     & 0.69                     & 0.44                     & 0.93                     & 0.92                     & 0.96                     & 0.95                    \\
\textbf{s5}  & 0.94                     & 0.75                     & 0.60                     & 0.63                     & 1.05                     & 1.06                     & 1.00                     & 1.03                     \\
\textbf{s6}  & 0.98                     & 0.86                     & 0.80                     & 0.63                     & 1.15                     & 1.60                     & 1.47                     & 1.11                     \\
\textbf{s7}  & 0.93                     & 0.81                     & 0.59                     & 0.62                     & 1.01                     & 1.12                     & 1.00                     & 0.97                     \\
\textbf{s8}  & 0.89                     & 0.62                     & 0.68                     & 0.59                     & 0.99                     & 1.18                     & 1.35                     & 1.11                     \\
\textbf{s9}  & 0.92                     & 0.81                     & 0.70                     & 0.54                     & 0.99                     & 0.99                     & 1.02                     & 0.90                     \\
\textbf{s10} & 0.97                     & 0.70                     & 0.58                     & 0.56                     & 1.05                     & 1.18                     & 1.13                     & 1.04                     \\
\textbf{s11} & 0.80                     & 0.73                     & 0.59                     & 0.50                     & 0.85                     & 0.89                     & 0.99                     & 0.96                     \\
\textbf{s12} & 0.80                     & 0.81                     & 0.64                     & 0.54                     & 1.01                     & 0.92                     & 1.01                     & 0.99                     \\
\textbf{s13} & 0.80                     & 0.75                     & 0.70                     & 0.75                     & 0.91                     & 1.30                     & 1.33                     & 1.03                     \\
\textbf{s14} & 0.91                     & 0.66                     & 0.57                     & 0.59                     & 1.08                     & 1.25                     & 1.05                     & 0.93                     \\
\textbf{s15} & 0.94                     & 0.78                     & 0.50                     & 0.57                     & 0.88                     & 0.92                     & 0.91                     & 0.99                     \\
\textbf{s16} & 0.94                     & 0.85                     & 0.73                     & 0.55                     & 0.88                     & 1.06                     & 1.04                     & 0.95                     \\
\textbf{s17} & 0.98                     & 0.83                     & 0.78                     & 0.59                     & 0.95                     & 1.04                     & 1.02                     & 1.11                     \\
\textbf{s18} & 0.92                     & 0.84                     & 0.69                     & -                         & 1.07                     & 1.01                     & 1.04                     & -                         \\
\textbf{s19} & 0.82                     & 0.79                     & 0.68                     & -                         & 0.94                     & 1.18                     & 1.04                     & -                         \\
\textbf{s20} & 0.90                     & 0.71                     & 0.68                     & -                         & 0.92                     & 1.23                     & 1.10                     & -                         \\
\textbf{s21} & 0.75                     & 0.69                     & 0.71                     & -                         & 0.74                     & 0.92                     & 0.92                     & -                         \\
\textbf{s22} & 0.95                     & 0.85                     & 0.72                     & -                         & 0.93                     & 0.98                     & 0.99                     & -                         \\
\textbf{s23} & 0.94                     & 0.77                     & 0.64                     & -                         & 0.98                     & 1.28                     & 1.20                     & -                         \\
\textbf{s24} & 1.00                     & 0.86                     & 0.72                     & -                         & 0.90                     & 1.00                     & 1.00                     & -                         \\ \hline 
\hline
\textbf{AVG}          & 0.90 & 0.77$^{*}$   & 0.67$^{*}$   & 0.58$^{*}$   & 0.95 & 1.07$^{*}$   & 1.06$^{*}$   & 1.00   \\
\textbf{STD}          & 0.07 & 0.07   & 0.07   & 0.06   & 0.09 & 0.17   & 0.14   & 0.06   \\ \hline
\end{tabular*}
\caption{AUC and SNR of ERP in scalp-EEG. Asterisk indicates significance levels of 5\% between the performance at 0\,m/s and corresponding speed.}
\label{tab:tab2}
\end{table*}

\begin{table*}[t!]
\centering
\small
\renewcommand{\arraystretch}{1.3}
\renewcommand{\tabcolsep}{5.05mm}
\begin{tabular*}{\textwidth}{c|c|c|c|c|c|c|c|c}
\hline
\textbf{Measure}      & \multicolumn{4}{c|}{\textbf{AUC}}    & \multicolumn{4}{c}{\textbf{SNR}}    \\ \hline
\textbf{Speed} & { 0 m/s } & {0.8 m/s} & {1.6 m/s} & {2.0 m/s} & { 0 m/s } & {0.8 m/s} & {1.6 m/s} & {2.0 m/s} \\ \hline
\textbf{s1}  & 0.52                     & 0.55                     & 0.52                     & 0.49                     & 1.28                     & 0.92                     & 1.06                     & 0.97                     \\
\textbf{s2}  & 0.71                     & 0.69                     & 0.52                     & 0.53                     & 0.93                     & 0.94                     & 0.97                     & 1.04                     \\
\textbf{s3}  & 0.66                     & 0.57                     & 0.57                     & 0.54                     & 0.77                     & 0.92                     & 0.86                     & 0.97                     \\
\textbf{s4}  & 0.54                     & 0.54                     & 0.53                     & 0.58                     & 1.05                     & 1.13                     & 0.98                     & 0.83                     \\
\textbf{s5}  & 0.94                     & 0.75                     & 0.63                     & 0.53                     & 0.97                     & 0.99                     & 1.01                     & 0.87                     \\
\textbf{s6}  & 0.86                     & 0.70                     & 0.65                     & 0.51                     & 1.20                     & 1.17                     & 1.09                     & 1.04                     \\
\textbf{s7}  & 0.85                     & 0.74                     & 0.57                     & 0.56                     & 1.06                     & 1.08                     & 1.01                     & 1.01                     \\
\textbf{s8}  & 0.57                     & 0.58                     & 0.59                     & 0.51                     & 1.20                     & 1.06                     & 1.03                     & 1.01                     \\
\textbf{s9}  & 0.81                     & 0.76                     & 0.70                     & 0.60                     & 0.93                     & 1.01                     & 1.01                     & 0.95                     \\
\textbf{s10} & 0.72                     & 0.64                     & 0.58                     & 0.49                     & 0.98                     & 0.98                     & 0.98                     & 0.92                     \\
\textbf{s11} & 0.69                     & 0.47                     & 0.55                     & 0.47                     & 1.07                     & 0.94                     & 0.96                     & 1.11                     \\
\textbf{s12} & 0.64                     & 0.60                     & 0.56                     & 0.54                     & 0.98                     & 0.94                     & 1.00                     & 0.97                     \\
\textbf{s13} & 0.46                     & 0.46                     & 0.62                     & 0.49                     & 0.97                     & 1.08                     & 1.06                     & 1.02                     \\
\textbf{s14} & 0.77                     & 0.58                     & 0.46                     & 0.49                     & 1.09                     & 1.09                     & 0.95                     & 0.97                     \\
\textbf{s15} & 0.76                     & 0.56                     & 0.51                     & 0.55                     & 0.97                     & 0.97                     & 0.94                     & 0.96                     \\
\textbf{s16} & 0.79                     & 0.71                     & 0.57                     & 0.51                     & 1.01                     & 1.06                     & 1.00                     & 0.99                     \\
\textbf{s17} & 0.87                     & 0.80                     & 0.67                     & 0.64                     & 1.01                     & 0.97                     & 1.00                     & 0.99                     \\
\textbf{s18} & 0.51                     & 0.68                     & 0.58                     & -                         & 0.99                     & 1.03                     & 0.99                     & -                         \\
\textbf{s19} & 0.72                     & 0.69                     & 0.73                     & -                         & 0.88                     & 1.24                     & 0.94                     & -                         \\
\textbf{s20} & 0.51                     & 0.41                     & 0.58                     & -                         & 2.24                     & 1.08                     & 0.88                     & -                         \\
\textbf{s21} & 0.69                     & 0.64                     & 0.52                     & -                         & 0.88                     & 1.01                     & 0.97                     & -                         \\
\textbf{s22} & 0.88                     & 0.68                     & 0.55                     & -                         & 0.95                     & 1.01                     & 0.98                     & -                         \\
\textbf{s23} & 0.80                     & 0.69                     & 0.56                     & -                         & 0.94                     & 0.97                     & 0.97                     & -                         \\
\textbf{s24} & 0.90                     & 0.73                     & 0.64                     & -                         & 0.95                     & 0.96                     & 0.96                     & -                         \\ \hline 
\hline
\textbf{AVG}    & 0.72 & 0.63$^{*}$   & 0.58$^{*}$   & 0.53$^{*}$   & 1.06 & 1.02   & 0.98   & 0.98   \\
\textbf{STD} & 0.14 & 0.10   & 0.06   & 0.04   & 0.27 & 0.08   & 0.05   & 0.06   \\ \hline
\end{tabular*}
\caption{AUC and SNR of ERP in ear-EEG. Asterisk indicates significance levels of 5\% between the performance at 0\,m/s and corresponding speed.}
\label{tab:tab3}
\end{table*}

\begin{table*}[t!]
\centering
\small
\renewcommand{\arraystretch}{1.3}
\renewcommand{\tabcolsep}{5.05mm}
\begin{tabular*}{\textwidth}{c|c|c|c|c|c|c|c|c}
\hline
\textbf{Measure}             & \multicolumn{4}{c|}{\textbf{Accuracy (\%)}}  & \multicolumn{4}{c}{\textbf{SNR}}  \\\hline
\textbf{Speed} & { 0 m/s } & {0.8 m/s} & {1.6 m/s} & {2.0 m/s} & { 0 m/s } & {0.8 m/s} & {1.6 m/s} & {2.0 m/s} \\ \hline
\textbf{s1}  & 100                   & 98.33                    & 98.33                    & 100                   & 4.60                     & 3.82                     & 2.27                     & 2.48                     \\
\textbf{s2}  & 100                   & 100                   & 98.33                    & 90.00                    & 3.93                     & 3.42                     & 3.01                     & 2.91                     \\
\textbf{s3}  & 96.67                    & 90.00                    & 83.33                    & 33.33                    & 1.52                     & 1.69                     & 1.35                     & 1.59                     \\
\textbf{s4}  & 100                   & 100                   & 100                   & 93.33                    & 2.43                     & 2.09                     & 2.06                     & 1.79                     \\
\textbf{s5}  & 96.67                    & 96.67                    & 96.67                    & 31.67                    & 3.79                     & 3.42                     & 3.42                     & 1.63                     \\
\textbf{s6}  & 100                   & 81.67                    & 75.00                    & 35.00                    & 2.35                     & 2.18                     & 1.88                     & 1.36                     \\
\textbf{s7}  & 100                   & 98.33                    & 96.67                    & 35.00                    & 4.12                     & 3.68                     & 2.87                     & 2.66                     \\
\textbf{s8}  & 96.67                    & 81.67                    & 65.00                    & 41.67                    & 1.14                     & 1.31                     & 1.01                     & 1.20                     \\
\textbf{s9}  & 98.33                    & 81.67                    & 98.33                    & 33.33                    & 2.46                     & 1.66                     & 1.90                     & 2.31                     \\
\textbf{s10} & 100                  & 86.67                    & 78.33                    & 35.00                    & 3.42                     & 1.88                     & 1.90                     & 1.74                     \\
\textbf{s11} & 100                   & 86.67                    & 81.67                    & 50.00                    & 2.19                     & 1.71                     & 1.99                     & 2.05                     \\
\textbf{s12} & 100                   & 96.67                    & 90.00                    & 60.00                    & 2.33                     & 2.16                     & 1.91                     & 1.45                     \\
\textbf{s13} & 80.00                    & 66.67                    & 80.00                    & 71.67                    & 1.32                     & 1.23                     & 1.11                     & 1.45                     \\
\textbf{s14} & 100                   & 96.67                    & 90.00                    & 36.17                    & 2.87                     & 2.55                     & 2.22                     & 2.19                     \\
\textbf{s15} & 100                   & 100                   & 100                   & 98.33                    & 2.87                     & 2.82                     & 2.39                     & 2.57                     \\
\textbf{s16} & 78.33                    & 93.33                    & 96.67                    & -                          & 2.72                     & 2.22                     & 2.09                     & -                         \\
\textbf{s17} & 80.00                    & 70.00                    & 56.67                    & -                          & 1.84                     & 1.35                     & 1.13                     & -                         \\
\textbf{s18} & 91.67                    & 63.33                    & 48.33                    & 31.67                    & 1.43                     & 1.32                     & 1.17                     & 1.63                     \\
\textbf{s19} & 33.33                    & 33.33                    & 31.67                    & -                          & 1.17                     & 1.15                     & 0.99                     & -                         \\
\textbf{s20} & 45.00                    & 48.33                    & 50.00                    & -                         & 1.32                     & 1.27                     & 1.19                     & -                         \\
\textbf{s21} & 96.67                    & 91.67                    & 98.33                    & -                         & 2.74                     & 2.60                     & 2.81                     & -                         \\
\textbf{s22} & 100                   & 100                   & 95.00                    & -                         & 2.81                     & 2.67                     & 2.29                     & -                         \\
\textbf{s23} & 46.67                     & 50.00                     & 46.67                     &   -                       & 1.24                     & 1.02                     & 1.13                     & -                         \\ \hline\hline
\textbf{AVG} & 88.70 & 83.12$^{*}$   & 80.65$^{*}$   & 54.76$^{*}$ & 2.64 & 2.14$^{*}$   & 1.92$^{*}$   & 1.94$^{*}$ \\
\textbf{STD} & 19.52 & 18.68   & 20.38   & 25.84 & 0.99 & 0.84   & 0.68   & 0.51 \\ \hline
\end{tabular*}
\caption{Accuracy and SNR of SSVEP in scalp-EEG. Asterisk indicates significance levels of 5\% between the performance at 0\,m/s and corresponding speed.}
\label{tab:tab4}
\end{table*}

\begin{table*}[t!]
\centering
\small
\renewcommand{\arraystretch}{1.3}
\renewcommand{\tabcolsep}{5.05mm}
\begin{tabular*}{\textwidth}{c|c|c|c|c|c|c|c|c}
\hline
\textbf{Measure}              & \multicolumn{4}{c|}{\textbf{Accuracy (\%)}}  & \multicolumn{4}{c}{\textbf{SNR}}  \\\hline
\textbf{Speed} & { 0 m/s } & {0.8 m/s} & {1.6 m/s} & {2.0 m/s} & { 0 m/s } & {0.8 m/s} & {1.6 m/s} & {2.0 m/s} \\ \hline
\textbf{s1}  & 96.67                     & 73.33                     & 46.67                     & 26.67                     & 2.06                     & 1.63                     & 1.32                     & 1.09                     \\
\textbf{s2}  & 61.67                     & 41.67                     & 36.67                     & 33.33                     & 1.37                     & 1.10                     & 0.86                     & 1.10                     \\
\textbf{s3}  & 53.33                     & 46.67                     & 36.67                     & 33.33                     & 1.15                     & 1.25                     & 0.97                     & 2.04                     \\
\textbf{s4}  & 55.00                     & 41.67                     & 33.33                     & 43.33                     & 1.11                     & 0.96                     & 0.94                     & 1.50                     \\
\textbf{s5}  & 68.33                     & 68.33                     & 53.33                     & 28.33                     & 1.57                     & 1.27                     & 1.08                     & 1.31                     \\
\textbf{s6}  & 41.67                     & 38.33                     & 40.00                     & 33.33                     & 1.04                     & 1.08                     & 1.07                     & 1.19                     \\
\textbf{s7}  & 58.33                     & 48.33                     & 41.67                     & 33.33                     & 1.18                     & 1.21                     & 1.11                     & 1.29                     \\
\textbf{s8}  & 58.33                     & 55.00                     & 41.67                     & 25.00                     & 1.08                     & 1.11                     & 1.14                     & 1.12                     \\
\textbf{s9}  & 60.00                     & 51.67                     & 38.33                     & 35.00                     & 1.17                     & 1.02                     & 0.92                     & 1.55                     \\
\textbf{s10} & 51.67                     & 40.00                     & 38.33                     & 31.67                     & 1.11                     & 1.06                     & 1.05                     & 1.14                     \\
\textbf{s11} & 35.00                     & 33.33                     & 36.67                     & 35.00                     & 1.07                     & 0.98                     & 0.95                     & 1.21                     \\
\textbf{s12} & 75.00                     & 55.00                     & 38.33                     & 36.67                     & 1.46                     & 1.40                     & 1.12                     & 1.10                     \\
\textbf{s13} & 46.67                     & 38.33                     & 50.00                     & 36.67                     & 1.09                     & 1.09                     & 1.03                     & 1.22                     \\
\textbf{s14} & 51.67                     & 41.67                     & 35.00                     & 36.17                     & 1.19                     & 0.94                     & 0.95                     & 1.24                     \\
\textbf{s15} & 50.00                     & 38.33                     & 35.00                     & 36.67                     & 1.42                     & 1.18                     & 0.97                     & 1.01                     \\
\textbf{s16} & 48.33                     & 48.33                     & 46.67                     & -                         & 1.14                     & 1.08                     & 1.06                     & -                        \\
\textbf{s17} & 46.67                     & 38.33                     & 35.00                     & -                         & 1.05                     & 0.93                     & 0.93                     & -                        \\
\textbf{s18} & 48.33                     & 45.00                     & 30.00                     & 28.33                     & 1.13                     & 0.87                     & 1.13                     & 1.31                     \\
\textbf{s19} & 40.00                     & 25.00                     & 35.00                     & -                         & 0.98                     & 0.89                     & 1.04                     & -                        \\
\textbf{s20} & 40.00                     & 30.00                     & 35.00                     & -                         & 1.02                     & 0.93                     & 0.93                     & -                        \\
\textbf{s21} & 63.33                     & 50.00                     & 55.00                     & -                         & 1.06                     & 1.14                     & 1.11                     & -                        \\
\textbf{s22} & 43.33                     & 50.00                     & 36.67                     & -                         & 1.27                     & 1.05                     & 1.01                     & -                        \\
\textbf{s23} & 30.00                     & 31.67                     & 35.00                     & -                         & 1.05                     & 1.00                     & 1.08                     & -                        \\ \hline\hline
\textbf{AVG} & 53.19 & 44.78$^{*}$   & 39.57$^{*}$   & 33.30$^{*}$ & 1.21 & 1.09$^{*}$   & 1.03$^{*}$   & 1.27 \\
\textbf{STD} & 13.93 & 11.11   & 6.39   & 4.43 & 0.23 & 0.17   & 0.10   & 0.24 \\ \hline
\end{tabular*}
\caption{Accuracy and SNR of SSVEP in ear-EEG. Asterisk indicates significance levels of 5\% between the performance at 0\,m/s and corresponding speed.}
\label{tab:tab5}
\end{table*}

\end{document}